%% file: workshop/workshop_paper.tex
\documentclass{article}

\usepackage[dblblindworkshop, final]{neurips_infpriv_2026}

\usepackage[utf8]{inputenc} 
\usepackage[T1]{fontenc}    
\usepackage{hyperref}       
\usepackage{url}            
\usepackage{booktabs}       
\usepackage{amsfonts}       
\usepackage{nicefrac}       
\usepackage{microtype}      
\usepackage{xcolor}         
\usepackage{tikz}
\usepackage{amsmath}
\usepackage{booktabs}
\usepackage{multirow}
\usetikzlibrary{positioning,arrows.meta,fit,backgrounds,calc,decorations.pathreplacing}
\usepackage{algorithm}
\usepackage{algpseudocodex}
\usepackage{amsmath,amssymb}
\usepackage{booktabs}
\usepackage{multirow}
\usepackage{soul}
\usepackage{subcaption}
\usepackage{pgfplots}

\usepackage[normalem]{ulem}

\definecolor{baseA}{RGB}{ 90,130,180}   
\definecolor{baseB}{RGB}{240,175, 90}   
\definecolor{ours} {RGB}{200, 45, 60}   
 
\title{
Membership Inference Attack on Private Evolution}

\author{%
  Daniil Filienko \\
  University of Washington Tacoma\\
  \texttt{daniilf@uw.edu} \\
  \And
  Sikha Pentyala \\
  University of Washington Tacoma \\
  \texttt{sikha@uw.edu} \\
  \AND
  Martine De Cock \\
  University of Washington Tacoma \\
  \texttt{mdecock@uw.edu} \\
}

\begin{document}

\maketitle

\begin{abstract}
Synthetic Data Generation (SDG) is an actively studied way to enable safe sharing of private data more broadly. Inference-only SDG methods, such as Private Evolution (PE), are particularly appealing, as they can leverage black-box access to pre-trained foundation models to generate high quality data without the need for model fine-tuning on the private data. PE is explicitly designed to produce synthetic data under Differential Privacy (DP). Yet formal DP guarantees do not eliminate the need for empirical privacy auditing of the practical susceptibility of released synthetic data to adversarial attacks. In this work, we introduce the first white-box Membership Inference Attack (MIA) tailored to the voting-based algorithm underlying the PE framework. By exploiting how the iterative filtering process biases the survival and proximity of synthetic candidates toward private instances, our attack achieves empirical performance indicating private data leakage in the PE applications across image, text, and tabular modalities.
    
\end{abstract}




\section{Introduction}
A lot of sensitive data cannot be easily shared because of privacy concerns. Synthetic Data Generation (SDG), i.e., the process of creating artificial data that looks like real data but without revealing sensitive information contained in it, is an actively studied way to operationalize such ``hidden'' data more broadly \cite{hu2023sok}. Traditional SDG methods rely on training a model on the real data, often with a Differential Privacy (DP) mechanism, and subsequently using the trained model to generate synthetic data. Since the emergence of foundational models, however, alternative ways of generating synthetic data have been investigated, since DP training or DP fine-tuning of foundation models is prohibitively computationally expensive. 
An appealing alternative are inference-only SDG methods such as Private Evolution (PE) that rely on API-only access to pre-trained foundation models. \textcolor{black}{At each iteration of PE, the real data holder prompts the model to generate a set of synthetic candidates followed by a voting process where private instances ``vote'' to retain the most similar synthetic candidates, such that the synthetic data distribution converges to the real distribution over $T$ rounds \cite{lin2024differentially}.} PE has been applied across text \cite{xiedifferentially, zhangpcevolve2025, zou2026pesgd, zou2025contrastive}, image \cite{lin2024differentially, wang2026synthesize}, and tabular data \cite{swanberg2025api,trandifferentially2026} in both centralized \cite{chien2026maple} and federated \cite{hou2024pre} settings.
Extending its influence beyond academic literature, PE-based methods have reached industrial deployment. In addition to Microsoft, Apple has recently deployed a PE-style voting mechanism in Apple Intelligence, where on-device embeddings of user data are matched against candidate synthetic embeddings under DP guarantees to identify aggregate trends for features such as email summarization, without ever accessing raw user data.\footnote{\url{https://machinelearning.apple.com/research/differential-privacy-aggregate-trends}}
These industry deployments mean that the privacy of a large number of people rests on the correctness of PE's theoretical guarantees.

\textcolor{black}{Empirical privacy auditing provides an important complementary perspective on the practical susceptibility of released synthetic data to adversarial attacks.}
The most established way to empirically verify the privacy of an SDG method is with membership inference attacks (MIA) \cite{shokri2017membership}, which detect whether specific private instances were used to train the synthetic data generation model \cite{golob2025privacy,hayes2017logan}. 

We propose a modality-agnostic MIA on PE, EVO-MIA, which uses the voting patterns of private data samples to determine whether a specific instance was used to shape the synthetic data distribution. It rests on the observation that in PE every private instance casts a vote for the synthetic sample most similar to it. Each private instance therefore leaves a signal, i.e., nudges up the vote count of its closest synthetic sample, while an instance that was never used leaves no such trace. As depicted in Figure~\ref{fig:voting_mia}, our attack only uses public hyperparameters and the DP-protected artifacts (noisy vote histograms) that the process releases during each iteration to determine whether for a given target instance $x^*$, the observed votes are better explained by that instance having participated in the voting process or not. 

Our attack is efficient and generalizable across modalities and can be applied to other voting-based SDG algorithms. It does not require training of shadow models, providing a practical way to verify the privacy protection of the algorithm before deployment. 


\input{diagram_overview}
\section{Methodology}
\paragraph{Formal problem description and threat model.}
We consider a data curator who holds a private dataset $D=\{x_1,x_2,\dots,x_n\}$ of $n$ instances (e.g.~texts, images, or records in a data table). The curator's goal is to release a synthetic corpus $D_\text{syn}$ with instances that look similar to $D$ without leaking sensitive information about individual instances in $D$.

To achieve this, PE iteratively utilizes a proprietary generative model $G$ and a public embedding model $\varphi$. In each round $t \in \{1, \dots, T\}$, $G$ proposes a set of synthetic candidate instances $D_\text{syn}^{(t)} = \{d^{(t)}_1, d^{(t)}_2, \dots, d^{(t)}_m\}$, where $m$ is the number of  synthetic instances to be generated.\footnote{Data curator chooses the value of $m$ and is passed as a parameter to PE. Specifically $m$ is set through setting hyperparameters $b$, the number of candidates selected each round, and $l$, the number of variations per selected candidate, so that $m = b \cdot l$, with $m$ split across classes in proportion to their frequencies in $D$.} The candidates $D_\text{syn}^{(T)}$ of the final round $T$ are released as the synthetic corpus $D_\text{syn}$.
In each round, PE uses a differentially private histogram mechanism where each private instance $x_i$ is mapped via $\varphi$ to an embedding and votes for its nearest synthetic candidate generated by $G$, such that more votes are given to the more realistic synthetic candidates. Votes $c$ are aggregated by the data curator and perturbed with Gaussian noise with standard deviation $\sigma$. A clipping threshold $\theta$ is then applied, subtracting $\theta$ from each noisy count, and negative results are released as $0$, i.e.~a candidate that received $\kappa$ votes has released count $c=\max(\kappa+\xi-\theta,0)$ with $\xi\sim\mathcal N(0,\sigma^2)$. Candidates with higher vote counts survive to be mutated by $G$ in the subsequent round. Since the private data
is accessed only through the noisy vote histograms, PE as an algorithm satisfies
$(\varepsilon,\delta)$-DP with respect to $D$.

For a target instance $x$, adversary $\mathcal{A}$ computes a score $s$ estimating whether $x\in D$. The adversary is honest-but-curious and has access to the public embedding model $\varphi$, the mechanism hyperparameters, i.e.~the clipping threshold $\theta$ and the noise standard deviation $\sigma$, which follows from the publicly disclosed privacy budget $(\varepsilon,\delta)$, the size $|D|$ of the private dataset, an in-distribution auxiliary set $\mathrm{Ref}$, and the per-round releases ${(D_\text{syn}^{(t)},c^{(t)})}_{t=1}^{T}$, where $c^{(t)}_j$ is the noisy vote count for each synthetic $d^{(t)}_j$.\footnote{Following Kerckhoffs's principle, we treat all mechanism parameters as public. The holder of the proprietary model $G$ constitutes a natural instantiation of this adversary in the centralized setting.}

This threat model applies to both federated and centralized settings with few modifications. In the \emph{federated} PE setup \cite{hou2024pre}, the noised vote histogram is broadcast to all participants to agree on which candidates to keep. This allows any participant or the aggregation server to log the full interaction transcript, including the counts $c^{(t)}$, which is the assumed setting for our EVO-MIA attack in Algorithm~\ref{alg:histogram-mia}. Alternatively, in the \emph{centralized} setup, the provider of the foundation model $G$ observes every candidate it is asked to mutate and which candidates survive each round, but does not observe the counts $c^{(t)}$ when the curator performs voting locally. For this weaker adversary, we propose a histogram-free variant, EVO-MIA-BB, which relies solely on candidate survival (Appendix~\ref{app:black_box_evo_mia})

We evaluate the attack on a target set $D_{\mathrm{target}} = M \cup \bar{M}$ of \emph{members} $M$ and \emph{non-members} $\bar{M}$, where members $M\subseteq D$ are instances used in the PE algorithm for voting, while non-members $\bar{M}$ are held-out instances drawn from the same distribution as the PE members. Attack quality is the ROC-AUC of the scores $s$ against the membership labels. This data split is depicted in Figure \ref{fig:data_universe}.

\begin{algorithm}[h]
\caption{EVO-MIA: Membership inference for a single target instance}
\label{alg:histogram-mia}
\begin{algorithmic}[1]
\Require target instance $x$; public embedding model $\varphi$; per-round releases $\{(D_\text{syn}^{(t)},c^{(t)})\}_{t=1}^{T}$; auxiliary set $\mathrm{Ref}$; number $n$ of private instances; noise standard deviation $\sigma$; clipping threshold $\theta$
\State $s \gets 0$
\For{$t = 1$ {to} $T$}
    \State $j \gets \arg\min_{j'} \lVert \varphi(x) - \varphi(d^{(t)}_{j'}) \rVert$
        \Comment{find a synthetic candidate most similar to $x$}
    \State $q_j \gets \dfrac{\lvert\{r\in\mathrm{Ref}: \text{$r$'s nearest candidate is }j\}\rvert}{\lvert\mathrm{Ref}\rvert}$
        \Comment{calculate the fraction of $\mathrm{Ref}$ that votes for $j$}
    \State $\mu_0 \gets n q_j$;\quad $\mu_1 \gets (n-1)\,q_j$ \Comment{expected votes for $j$}
    \For{$\kappa = 0, 1, 2, \dots$}
    
        \If{counts are clipped {and} $c^{(t)}_j = 0$}
            \State $w_\kappa \gets \Phi\big((\theta-\kappa)/\sigma\big)$ \Comment{noisy count was below $\theta$}
        \Else
            \State $w_\kappa \gets \mathcal N\big(c^{(t)}_j+\theta;\,\kappa,\sigma^2\big)$ 
        \EndIf
    \EndFor
    \State $P_0 \gets \sum_{\kappa=0}^{\infty} w_\kappa\,\mathrm{NB}(\kappa\mid\mu_0)$ \Comment{probability of $c^{(t)}_j$ under $H_0$}
    \State $P_1 \gets \sum_{\kappa=1}^{\infty} w_\kappa\,\mathrm{NB}(\kappa-1\mid\mu_1)$ \Comment{probability of $c^{(t)}_j$ under $H_1$}
    \State $s \gets s + \log P_1 - \log P_0$

\EndFor
\State \Return $s$
\end{algorithmic}
\end{algorithm}

\paragraph{EVO-MIA.}
The insights underlying our ``membership inference via evolutionary traces'' attack (EVO-MIA) are based on the observation that PE is fundamentally an iterative filtering method. Each
PE step aims to discard synthetic candidates that are far from the private data, so that the final dataset keeps the synthetic data that is close to private instances. Thus, an adversary with access to the per-round vote counts can estimate the membership probability of the target instance $x$. For each evolution round, the attack identifies the synthetic candidate closest to the target instance $x$. It then uses the auxiliary reference set to estimate the expected vote distribution, computing a log-likelihood ratio that compares the probability of the observed noisy vote count if the instance were a member (hypothesis $H_1$) versus a non-member (hypothesis $H_0$). This ratio is accumulated across all rounds to produce the final membership score $s$ for $x$. A more formal description of our method can be found in Algorithm \ref{alg:histogram-mia}.

Following PE, which runs separately for each class, we run Algorithm~\ref{alg:histogram-mia} separately for each class. In each round, the adversary finds the candidate $j$ closest to $x$ (Line 3), estimates the fraction $q_j$ of instances voting for $j$ from $\mathrm{Ref}$ (Line 4), and computes the expected number of votes for $j$ if $x$ is a non-member or a member (Line 5). Since only the noisy, clipped count is released, on Lines 6--10 the adversary computes the probability $w_\kappa$ of the released count for every possible true count $\kappa$ under the observation model defined by PE's Gaussian mechanism , where $\mathcal N\big(c^{(t)}_j+\theta;\,\kappa,\sigma^2\big)$ is the Gaussian density and $\Phi$ the standard normal cumulative distribution function. Lines 11--12 weight $w_\kappa$ by the distribution of $\kappa$ under each hypothesis, and Line 13 adds the log-likelihood ratio to the score $s$.

The exact distribution of the true count is binomial, $\kappa\sim\mathrm{Bin}(n,q_j)$ under $H_0$, which is close to a Poisson distribution when $q_j$ is small. However, in our experiments the observed vote counts vary $1.5$--$3.8$ times more than a Poisson distribution allows, partly because $q_j$ is estimated from $\mathrm{Ref}$. We therefore use a negative binomial distribution $\mathrm{NB}(\cdot\mid\mu)$ with mean $\mu$ and variance $1.8\,\mu$ in all experiments. Summing the log-likelihood ratios over rounds treats the rounds as independent. This is exact for the noise, which is drawn independently in every round, but only an approximation for the underlying counts, which are correlated because the same private instances vote in every round. 

\section{Experiments and Results}
We evaluate EVO-MIA across text, image, and tabular modalities. Dataset details and the baseline MIA algorithms we compare against are described in Appendices~\ref{app:datasets} and~\ref{app:baseline_attacks} respectively. The average AUC of 3 runs are summarized in Table~\ref{tab:results_tab_short}, Table~\ref{tab:results_images}, and Table~\ref{tab:results_text}. \footnote{For computing attack AUC, we subsample the dataset to at most 2,000 members and 2,000 non-members. For datasets with fewer non-members, we use all available ones. Since both PE and our method generate and attack instances one class at a time, we pool all predictions across classes together to compute attack AUC.}


EVO-MIA performs strongest on text, achieving an AUC of $0.873$ at $\epsilon{=}\infty$ and $0.685$ at $\epsilon{=}10$ on OpenReview, and $0.758$ and $0.605$ respectively on MIMIC-III (Table~\ref{tab:results_text}), consistently exceeding the n-gram baseline \cite{sun2025synbench} across all $\epsilon$ values. Among images, Cat stands out with AUC $0.846$ at $\varepsilon{=}\infty$ and $0.692$ at $\varepsilon{=}10$ (Table~\ref{tab:results_images}), substantially above the LOGAN \cite{hayes2017logan} baseline. We hypothesize this performance is due to the high-dimensional nature of the generated synthetic data for text and image data, leading to more distinct histogram patterns for member points. For the tabular data (Table~\ref{tab:results_tab_short}), the performance seems to be on par with the baselines, slightly surpassing the existing state-of-the-art on the breast-cancer dataset in the high-epsilon setting. On adult and person-activity, all attacks remain close to chance at every $\epsilon$. These are the two largest datasets. We hypothesize that a single member's additional vote is small relative to the expected count of its nearest cell, so the per-round membership signal is weaker.

Notably, it appears that low utility does not imply low leakage. For instance at $\epsilon{=}10$,  synthetic text data falls below the the majority baseline in utility yet retains an AUC of $0.685$ (more utility vs.~ attack results are in Appendix~\ref{app:results_ext}).


\begin{table}[h]
\centering
\caption{{MIA results for PE-Tabular}}
\label{tab:results_tab_short}
\small
\begin{tabular}{llccc}
\toprule
 & & \multicolumn{3}{c}{AUC by privacy budget $\epsilon$} \\
\cmidrule(lr){3-5}
Dataset & Method & $\infty$ & 10 & 1 \\
\midrule
\multirow{3}{*}{adult}
 & GAN-Leaks       & $0.504\pm0.007$ & $0.505\pm0.005$ & $0.505\pm0.001$ \\
 & DOMIAS          & $0.509\pm0.007$ & $0.513\pm0.006$ & $0.501\pm0.013$ \\
 & EVO-MIA (ours)  & $0.529\pm0.006$ & $0.516\pm0.005$ & $0.506\pm0.012$ \\
\midrule
\multirow{3}{*}{artificial-characters}
 & GAN-Leaks       & $0.614\pm0.013$ & $0.605\pm0.006$ & $0.572\pm0.018$ \\
 & DOMIAS          & $0.556\pm0.024$ & $0.558\pm0.033$ & $0.556\pm0.024$ \\
 & EVO-MIA (ours)  & $0.631\pm0.023$ & $0.604\pm0.022$ & $0.545\pm0.028$ \\
\midrule
\multirow{3}{*}{breast-cancer}
 & GAN-Leaks       & $0.530\pm0.009$ & $0.493\pm0.010$ & $0.554\pm0.015$ \\
 & DOMIAS          & $0.553\pm0.020$ & $0.473\pm0.015$ & $0.538\pm0.025$ \\
 & EVO-MIA (ours)  & $0.655\pm0.017$ & $0.573\pm0.042$ & $0.538\pm0.014$ \\
\midrule
\multirow{3}{*}{person-activity}
 & GAN-Leaks       & $0.520\pm0.008$ & $0.519\pm0.004$ & $0.507\pm0.008$ \\
 & DOMIAS          & $0.509\pm0.013$ & $0.507\pm0.010$ & $0.502\pm0.002$ \\
 & EVO-MIA (ours)  & $0.507\pm0.001$ & $0.509\pm0.011$ & $0.503\pm0.006$ \\
\bottomrule
\end{tabular}
\end{table}

\begin{table}[h]
\centering
\caption{{MIA results for PE-Images}}
\label{tab:results_images}
\small
\begin{tabular}{llccc}
\toprule
 & & \multicolumn{3}{c}{AUC by privacy budget $\epsilon$} \\
\cmidrule(lr){3-5}
Dataset & Method & $\infty$ & 10 & 1 \\
\midrule
\multirow{2}{*}{CIFAR-10}
 & LOGAN           & $0.508\pm0.003$ & $0.508\pm0.002$ & $0.508\pm0.003$ \\
 & EVO-MIA (ours)  & $0.696\pm0.005$ & $0.522\pm0.003$ & $0.534\pm0.006$ \\
\midrule
\multirow{2}{*}{Camelyon17}
 & LOGAN           & $0.513\pm0.007$ & $0.513\pm0.007$ & $0.513\pm0.007$ \\
 & EVO-MIA (ours)  & $0.638\pm0.005$ & $0.559\pm0.006$ & $0.524\pm0.005$ \\
\midrule
\multirow{2}{*}{Cat}
 & LOGAN           & $0.507\pm0.001$ & $0.507\pm0.001$ & $0.508\pm0.000$ \\
 & EVO-MIA (ours)  & $0.846\pm0.000$ & $0.692\pm0.000$ & $0.585\pm0.000$ \\
\bottomrule
\end{tabular}
\end{table}

\begin{table}[!h]
\centering
\caption{{MIA results for PE-text}}
\label{tab:results_text}
\small
\begin{tabular}{llccc}
\toprule
 & & \multicolumn{3}{c}{AUC by privacy budget $\epsilon$} \\
\cmidrule(lr){3-5}
Dataset & MIA & $\infty$ & 10 & 1 \\
\midrule
\multirow{2}{*}{OpenReview}
 & EVO-MIA (ours)                    & $0.873\pm0.004$ & $0.685\pm0.008$ & $0.551\pm0.006$ \\
 & n-gram MIA \cite{sun2025synbench} & $0.490\pm0.000$ & $0.491\pm0.000$ & $0.493\pm0.000$ \\
\midrule
\multirow{2}{*}{MIMIC-III}
 & EVO-MIA (ours)                    & $0.758\pm0.001$ & $0.605\pm0.002$ & $0.546\pm0.004$ \\
 & n-gram MIA \cite{sun2025synbench} & $0.496\pm0.001$ & $0.494\pm0.001$ & $0.493\pm0.001$ \\
\bottomrule
\end{tabular}
\end{table}





\section{Conclusion and Future Work}
We introduced a first membership inference attack on PE-based framework working across different modalities, designed to fill in the existing research gap in the MIAs developed for inference-only synthetic data generators.

In future work, we plan to examine tighter privacy auditing  \cite{auditing2023tight}, exploiting the fact that the DP step in PE is a simple analytical function, meaning that the privacy loss of the voting mechanism can be evaluated in closed form for known datasets, which makes tight privacy auditing for PE considerably easier than for DP-SGD. We also intend to further improve this attack using novel LLM-based methods which has been shown to be effective in discovering stronger MIAs \cite{tran2026automated}.

\section*{Limitation}
We evaluate our method on the datasets published in the original works, which produces results that may not perfectly generalize to other settings. 

Our attack relies on access to the per-round synthetic candidates and, in its main form, the noisy vote histograms. Our threat model therefore assumes that per-round releases are observable. Deployments that keep intermediate rounds private would reduce this exposure. Our histogram-free variant, EVO-MIA-BB (Appendix~\ref{app:black_box_evo_mia}), relaxes the need for vote counts but still requires per-round candidates.

\section*{Acknowledgments}
This material is based upon work
supported by the National Science Foundation under Grant No.~2451163 and No.~2523406. 

We thank reviewers of InfPriv workshop at NeurIPS 2026 for their detailed feedback.

\newpage
\bibliographystyle{plain}
\bibliography{references}


\newpage
\appendix

\section{Datasets}
\label{app:datasets}

Figure~\ref{fig:data_universe} illustrates the data setup and the notations used for various subsets of the data.

\input{diagram_setup}

We report attack AUC on  $D_{\mathrm{target}}$ on the datasets used in the original PE works for images \cite{lin2024differentially} and tabular \cite{trandifferentially2026} modalities. For text, we also use the original OpenReview dataset \cite{xiedifferentially}, as well as  a MIMIC-based dataset \cite{eckardt2023mimicking} to ensure that our findings generalize to realistic confidential data \cite{sun2025evaluating}. The \emph{members} are the private records used to vote for the synthetic samples, and the \emph{non-members} are a disjoint set of in-distribution records the generator never saw, drawn from the same dataset. The attack additionally uses a third, disjoint reference set $\mathrm{Ref}$. This is an in-distribution dataset drawn from the original dataset and never used for voting and separate from the scored non-members.

\begin{table}[h]
\centering
\caption{Datasets used in our evaluation, grouped by modality, with train subset from which members are samples, test set which constitutes non-members, and reference set sizes}
\label{tab:datasets}
\begin{tabular}{llrrr}
\toprule
Modality & Dataset & $D_{\text{train}}$ & $D_{\text{test}}$ & $n_{\text{Ref}}$ \\
\midrule
\multirow{2}{*}{Text}
 & OpenReview            & 8{,}396   & 2{,}797  & 2{,}798 \\
 & MIMIC-III             & 7{,}186   & 859      & 899     \\
\midrule
\multirow{3}{*}{Images}
 & CIFAR-10              & 10{,}000  & 2{,}000  & 2{,}000 \\
 & Camelyon17            & 4{,}000   & 2{,}000  & 2{,}000 \\
 & Cat (AFHQ)            & 136       & 34       & 30      \\
\midrule
\multirow{4}{*}{Tabular}
 & adult                 & 33{,}334  & 3{,}567  & 2{,}876 \\
 & artificial-characters & 7{,}152   & 226      & 766     \\
 & breast-cancer         & 398       & 43       & 43      \\
 & person-activity       & 115{,}402 & 12{,}365 & 9{,}763 \\
\bottomrule
\end{tabular}
\end{table}

\paragraph{Text.} OpenReview consists of ICLR'23 peer reviews, conditioned on joint area $\times$ rating classes. MIMIC-III consists of hospital discharge notes, conditioned on 5 diagnosis classes.

\paragraph{Images.} CIFAR-10 comprises 32px natural images across 10 classes. Camelyon17 comprises 64px histopathology patches (tumor/normal, 2 classes). Cat (AFHQ) comprises 512px cat faces across 2 label-ids.

\paragraph{Tabular.} adult, artificial-characters, breast-cancer, and person-activity are standard tabular benchmarks with utility target labels of income (2 classes), Class (10 classes), diagnosis (2 classes), and Class (11 classes), respectively.


\section{Baseline Attacks}
\label{app:baseline_attacks}
We compare our PE-specific MIA to state-of-the-art black-box MIA attacks across the modalities. For text-based Aug-PE \cite{xiedifferentially}, we use a recently proposed black-box n-gram based MIA \cite{sun2025synbench} which involves training an n-gram model on synthetic and reference datasets and using the difference in the output probability as the membership signal. For tabular data, we compare our method against the GAN-leaks attack \cite{chen2020gan}, which has been extensively validated in recent synthetic tabular data competitions \cite{menziessynthetic__camda,ozturk2026towards} and the more recent DOMIAS attack \cite{van2023membership}. And for images, we compare with the LOGAN attack \cite{hayes2017logan}, which trains a shadow  Generative Adversarial Network (GAN) model on the synthetic data and uses it as a membership classifier.

\section{Distribution Effect on MIA Performance}
\label{app:distribuion}
We also investigate the performance of the proposed attack on the artificial-characters dataset as related to the distribution shift between the members and auxiliary data with the results presented in Table \ref{tab:distribution}.

\paragraph{Split construction.}
We use the full artificial-characters dataset and project every row onto the first principal component (PCA-1). For split q$X$, $X\in\{50,35,25,18\}$, within each class we take the lowest $X\%$ of rows by PCA-1 score as the member pool and the highest $X\%$ as the auxiliary pool, and subsample each to 1,500 rows (class-stratified). Smaller $X$ means stronger shift, q50 splits each class into two adjacent halves, while q18 leaves the middle 64\% of the distribution between members and auxiliary data. The iid split is a class-stratified random disjoint split of the same pool. PE is run on the member pool only.

\begin{table}[htbp]
\centering
\caption{Distribution shift and privacy metrics across splits.}
\label{tab:distribution}
\begin{tabular}{lrrrr}
\toprule
Shift & embedding\_gap & pca1\_wasserstein & MIA\_AUC & utility (\%) \\
\midrule
iid & 0.015 & 0.053 & 0.602 & 47.4 \\
q50 & 0.420 & 2.054 & 0.979 & 54.7 \\
q35 & 0.567 & 2.768 & 0.992 & 36.8 \\
q25 & 0.677 & 3.294 & 0.997 & 33.4 \\
q18 & 0.775 & 3.769 & 1.000 & 21.1 \\
\bottomrule
\end{tabular}
\end{table}

\paragraph{Utility.}
Utility is measured by training a TabICL classifier \cite{qu2025tabicl} on the synthetic data and evaluating it on the real held-out test set, reporting the resulting accuracy on the 10-class classification task.

\paragraph{embedding\_gap.}
Computed as the L2 distance between the mean embedding of the members and the mean embedding of the auxiliary pool, using the full 7-feature tabular embedding space. This is a measure of separation between the two pools: values near zero (e.g., iid $=0.015$) indicate the datasets are very similar, while larger values (e.g., q18 $=0.775$) indicate the data is very different.

\paragraph{pca1\_wasserstein.}
The Wasserstein distance between the member and auxiliary datasets after applying Principal Component Analysis (PCA), denoting the separation between the datasets, similar to embedding\_gap.

\paragraph{MIA\_AUC.}
For the distribution experiments, $|D_{target}|=1,000$: 500 members drawn from the 1{,}501-row member pool, and 500 challenge non-members. We also reserve 750 rows as the reference set $D_{\text{ref}}$. 

In Table~\ref{tab:distribution}, non-members are sampled from the opposite tail of the distribution to members. The near-perfect AUCs therefore reflect this distribution difference, such that only the iid setting measures leakage with exchangeable members and non-members, showing the importance of members and non-members being drawn from the same distribution.

\section{Black-Box EVO-MIA}
\label{app:black_box_evo_mia}
In Algorithm \ref{alg:selection-mia} we describe an alternative attack using only the synthetic candidate survivorship captured in the set $\mathrm{Sel}^{(t)}$ of indices of the candidates that survived round $t$. The results of the attack are reported in Table \ref{tab:selection-mia-results}.


Text and tabular PE keep the candidates with the highest released counts, i.e.~those above a cut-off $\tau^{(t)}>0$, which the adversary has to estimate. A candidate with expected count $\mu$ survives a cut-off $\tau$ with probability
\[
P_\text{surv}(\tau\mid\mu)=\sum_{\kappa=0}^{\infty}\bar\Phi\big((\tau+\theta-\kappa)/\sigma\big)\,\mathrm{NB}(\kappa\mid\mu),
\]
where $\bar\Phi=1-\Phi$. The adversary first computes the expected vote count $\mu_{0,k}=n\,q_k$ of every candidate $k$ from $\mathrm{Ref}$ (Lines 6--7). This gives $f(\tau)=\sum_k P_\text{surv}(\tau\mid\mu_{0,k})$, the number of candidates expected to survive under $H_0$ if the cut-off were $\tau$. On Line 8, the adversary then searches by bisection for the $\tau$ at which $f(\tau)$ equals the observed number of survivors $V$. For example, suppose that $V=25$ of $100$ candidates survived. If $f(10)=40$, more candidates would survive than observed, so the cut-off must be higher. If $f(15)=20$, too few would survive, so the cut-off lies between $10$ and $15$, and the adversary next tries $12.5$, repeating this halving until $f(\tau)=V$. Lines 10--11 give the probability that $x$'s nearest candidate $j$ survives under each hypothesis, and Lines 13--16 add the log-likelihood ratio of whether $j$ survived to the score.

\begin{algorithm}[h]
\caption{EVO-MIA-BB: Black-Box Membership inference based on synthetic candidate survival}
\label{alg:selection-mia}
\begin{algorithmic}[1]
\Require target instance $x$; public embedding model $\varphi$; per-round releases
  $\{(D_\text{syn}^{(t)},\mathrm{Sel}^{(t)})\}_{t=1}^{T}$, where $D_\text{syn}^{(t)}=\{d^{(t)}_1,\dots,d^{(t)}_{m}\}$
  is round-$t$'s candidate pool and $\mathrm{Sel}^{(t)}\subseteq\{1,\dots,m\}$ is the
  index set of the candidates that survived into round $t+1$; auxiliary set
  $\mathrm{Ref}$; number $n$ of private instances; noise standard deviation $\sigma$; clipping threshold $\theta$
\Statex
\State $s \gets 0$
\For{$t = 1$ {to} $T$}
    \State $j \gets \arg\min_{j'} \lVert \varphi(x) - \varphi(d^{(t)}_{j'}) \rVert$ \Comment{find a synthetic candidate most similar to $x$}
    \State $V \gets \lvert \mathrm{Sel}^{(t)} \rvert$ \Comment{number of surviving candidates}
    \If{$V = 0$ {or} $V = m$}\ {continue} 
    
    \EndIf
    \For{$k = 1$ {to} $m$} \Comment{expected votes for every candidate under $H_0$}
        \State $q_{k} \gets \dfrac{\lvert\{r\in\mathrm{Ref}:\text{$r$'s nearest candidate is }k\}\rvert}{\lvert\mathrm{Ref}\rvert}$;
               \quad $\mu_{0,k} \gets n q_{k}$
    \EndFor
    \State $\tau^{(t)} \gets$ solve $\displaystyle\sum_{k=1}^{m} P_\text{surv}\big(\tau \mid \mu_{0,k}\big) = V$
        for $\tau$ by bisection 
    \State $\mu_0 \gets n q_j$;\quad $\mu_1 \gets (n-1)\,q_j$
    \State $P_0 \gets P_\text{surv}(\tau^{(t)} \mid \mu_0)$ \Comment{survival probability of $j$ under $H_0$}
    \State $\displaystyle P_1 \gets \sum_{\kappa=1}^{\infty}
        \bar\Phi\!\left(\frac{\tau^{(t)}+\theta-\kappa}{\sigma}\right)\,\mathrm{NB}(\kappa-1\mid\mu_1)$ \Comment{survival probability of $j$ under $H_1$}
    \State $z \gets \mathbf{1}[\,j \in \mathrm{Sel}^{(t)}\,]$ 
    \If{$z = 1$}
        \State $s \gets s + \log P_1 - \log P_0$
    \Else
        \State $s \gets s + \log(1-P_1) - \log(1-P_0)$
    \EndIf
\EndFor
\State \Return $s$
\end{algorithmic}
\end{algorithm}

Algorithm~\ref{alg:selection-mia} describes text and tabular PE, which keep the top-ranked candidates. Image PE instead samples the next round's candidates with replacement, in proportion to their noisy counts. For images we therefore replace the threshold step and the survival probabilities by the probability that candidate $j$ is drawn at least once. This is computed from the expected count of $j$ under each hypothesis, divided by the expected total over all candidates. The rest of the algorithm is not changed.

\begin{table}[h]
\centering
\caption{Performance results based on whether the noisy histogram counts are available (EVO-MIA) or not (EVO-MIA-BB) on datasets with $\epsilon=10$}
\label{tab:selection-mia-results}
\begin{tabular}{|l|l|c|c|}
\hline
{Modality} & {Dataset} & {EVO-MIA-BB} & {EVO-MIA} \\ \hline
Text              & OpenReview             & 0.650               & 0.671            \\ \hline
Tabular           & adult                  & 0.502               & 0.520            \\ \hline
Tabular           & artificial-characters  & 0.541               & 0.579            \\ \hline
Tabular           & breast-cancer          & 0.563               & 0.622            \\ \hline
Tabular           & person-activity        & 0.491               & 0.506            \\ \hline
Images            & Cat                    & 0.580               & 0.692            \\ \hline
Images            & CIFAR-10               & 0.497               & 0.528            \\ \hline
Images            & Camelyon17             & 0.505               & 0.555            \\ \hline
\end{tabular}
\end{table}

\section{Additional utility and MIA Results}
\label{app:results_ext}

\begin{table}[h]
\centering
\caption{{Results for PE-Tabular}}
\label{tab:results_tab}
\small
\begin{tabular}{llccccccc}
\toprule
 & & \multicolumn{7}{c}{AUC by privacy budget $\epsilon$} \\
\cmidrule(lr){3-9}
Dataset & Method & $\infty$ & 100 & 10 & 5 & 1 & 0.5 & 0.25 \\
\midrule
\multirow{3}{*}{adult}
 & GAN-Leaks       & 0.532 & 0.532 & 0.531 & 0.517 & 0.530 & 0.525 & 0.519 \\
 & DOMIAS          & 0.539 & 0.530 & 0.531 & 0.511 & 0.519 & 0.509 & 0.496 \\
 & EVO-MIA (ours)  & 0.523 & 0.519 & 0.517 & 0.515 & 0.492 & 0.487 & 0.472 \\
\midrule
\multirow{3}{*}{artificial-characters}
 & GAN-Leaks       & 0.616 & 0.585 & 0.605 & 0.580 & 0.561 & 0.528 & 0.552 \\
 & DOMIAS          & 0.537 & 0.531 & 0.518 & 0.549 & 0.519 & 0.475 & 0.511 \\
 & EVO-MIA (ours)  & 0.618 & 0.597 & 0.588 & 0.574 & 0.531 & 0.533 & 0.515 \\
\midrule
\multirow{3}{*}{breast-cancer}
 & GAN-Leaks       & 0.522 & 0.519 & 0.498 & 0.521 & 0.562 & 0.464 & 0.550 \\
 & DOMIAS          & 0.576 & 0.535 & 0.489 & 0.519 & 0.567 & 0.477 & 0.504 \\
 & EVO-MIA (ours)  & 0.675 & 0.551 & 0.622 & 0.522 & 0.528 & 0.492 & 0.562 \\
\midrule
\multirow{3}{*}{person-activity}
 & GAN-Leaks       & 0.510 & 0.494 & 0.482 & 0.492 & 0.485 & 0.479 & 0.491 \\
 & DOMIAS          & 0.524 & 0.516 & 0.508 & 0.515 & 0.485 & 0.496 & 0.523 \\
 & EVO-MIA (ours)  & 0.512 & 0.509 & 0.494 & 0.500 & 0.489 & 0.504 & 0.484 \\
\bottomrule
\end{tabular}
\end{table}

\begin{table}[h]
\centering
\caption{{PE-Tabular: MIA leakage vs.\ utility vs.\ fidelity}}
\label{tab:combined_tabular}
\small
\begin{tabular}{llccccccc}
\toprule
 & & \multicolumn{7}{c}{Value by privacy budget $\epsilon$} \\
\cmidrule(lr){3-9}
Dataset & Metric & $\infty$ & 100 & 10 & 5 & 1 & 0.5 & 0.25 \\
\midrule
\multirow{5}{*}{adult}
 & EVO-MIA AUC        & 0.523 & 0.519 & 0.517 & 0.515 & 0.492 & 0.487 & 0.472 \\
 & Utility Acc (\%)   & 80.6  & 81.5  & 82.3  & 81.0  & 79.7  & 81.0  & 80.9  \\
 & Fidelity (FID)     & 0.038 & 0.046 & 0.043 & 0.042 & 0.039 & 0.047 & 0.075 \\
 & Majority baseline (\%)  & \multicolumn{7}{c}{75.8} \\
 & Real data (Acc / FID)   & \multicolumn{7}{c}{86.5 / 0.0013} \\
\midrule
\multirow{5}{*}{artificial-characters}
 & EVO-MIA AUC        & 0.618 & 0.597 & 0.588 & 0.574 & 0.531 & 0.533 & 0.515 \\
 & Utility Acc (\%)   & 62.0  & 62.2  & 60.8  & 58.5  & 50.5  & 40.3  & 31.3  \\
 & Fidelity (FID)     & 0.0004 & 0.0006 & 0.0007 & 0.0015 & 0.0078 & 0.0258 & 0.1216 \\
 & Majority baseline (\%)  & \multicolumn{7}{c}{13.8} \\
 & Real data (Acc / FID)   & \multicolumn{7}{c}{90.4 / 0.0007} \\
\midrule
\multirow{5}{*}{breast-cancer}
 & EVO-MIA AUC        & 0.675 & 0.551 & 0.622 & 0.522 & 0.528 & 0.492 & 0.562 \\
 & Utility Acc (\%)   & 82.6  & 86.1  & 94.2  & 100.0 & 76.7  & 87.2  & 67.4  \\
 & Fidelity (FID)     & 0.440 & 0.595 & 0.397 & 1.008 & 2.132 & 2.678 & 3.012 \\
 & Majority baseline (\%)  & \multicolumn{7}{c}{62.8} \\
 & Real data (Acc / FID)   & \multicolumn{7}{c}{100.0 / 0.0276} \\
\midrule
\multirow{5}{*}{person-activity}
 & EVO-MIA AUC        & 0.512 & 0.509 & 0.494 & 0.500 & 0.489 & 0.504 & 0.484 \\
 & Utility Acc (\%)   & 68.5  & 69.4  & 69.1  & 67.9  & 63.1  & 59.9  & 55.9  \\
 & Fidelity (FID)     & 0.0011 & 0.0010 & 0.0011 & 0.0012 & 0.0034 & 0.0088 & 0.0248 \\
 & Majority baseline (\%)  & \multicolumn{7}{c}{33.0} \\
 & Real data (Acc / FID)   & \multicolumn{7}{c}{92.9 / 0.00004} \\
\bottomrule
\end{tabular}
\\[2pt]
\end{table}

\begin{table}[h]
\centering
\caption{{PE-Images ($\epsilon=10$): MIA leakage vs.\ utility vs.\ fidelity}}
\label{tab:combined_images}
\small
\begin{tabular}{lccccc}
\toprule
Dataset & EVO-MIA AUC & Synth. Acc (\%) / FID & Maj. baseline & Real data Acc (\%) / FID \\
\midrule
CIFAR-10 (3 rounds)   & 0.528 & 28.2 / 16.74 & 10.0\% & 27.2 / 16.60 \\
CIFAR-10 (10 rounds)  & 0.550 & 21.3 / 22.85 & 10.0\% & 30.0 / 16.60 \\
Camelyon17            & 0.555 & 46.3 / 95.66 & 50.0\% & 82.3 / 8.28 \\
Cat                   & 0.692 & n/a / 73.74  & n/a    & n/a / n/a \\
\bottomrule
\end{tabular}
\\[2pt]
\end{table}

\begin{table}[h]
\centering
\caption{{Performance of PE-Text on OpenReview, $\epsilon=10$}}
\label{tab:combined_text}
\small
\begin{tabular}{lcc}
\toprule
Metric & Synthetic & Real data \\
\midrule
EVO-MIA AUC          & 0.671 & --- \\
Utility Acc (\%)     & 26.3  & 30.0 \\
Majority baseline (\%) & \multicolumn{2}{c}{32.0} \\
FID                  & 31.3  & 2.83 \\
Precision            & 0.600 & 0.772 \\
Recall               & 0.167 & 0.729 \\
F1                   & 0.261 & 0.750 \\
\bottomrule
\end{tabular}
\\[2pt]
\end{table}

\section{Ablation Study Results}
\label{app:ablation}
We examine how the hyperparameters of PE and of the attack affect privacy leakage. All AUCs are for EVO-MIA (Algorithm~\ref{alg:histogram-mia}).

\paragraph{Number of evolution rounds.}
We vary the number of rounds $T$ in two ways. First, we rerun PE with $T\in\{1,3,6,9\}$ at a fixed privacy budget, so that more rounds require more noise per round. We compute attack AUC and utility which is TabICL \cite{qu2025tabicl} classification accuracy (Table~\ref{tab:abl-rounds-fixed}). Leakage grows with $T$ and saturates around $T\approx6$: more rounds give the adversary more observations of each target, but at a fixed $\varepsilon$ each observation becomes noisier. Utility keeps improving with $T$. Second, we attack only the first $T$ rounds of a single run, where the noise per round is fixed (Table~\ref{tab:abl-rounds-prefix}). Image and tabular runs have $10$ attackable rounds and text has $9$. Tabular PE requires $15$ iterations, but the first $5$ select candidates by sampling and release no vote count. Attack AUC grows steadily for text and images, while for tabular data it stays flat. 

\begin{table}[h]
\centering
\caption{AUC when PE is rerun with $T$ rounds at a fixed privacy budget on artificial-characters dataset. The noise $\sigma$ is recomputed for every $T$. }
\label{tab:abl-rounds-fixed}
\begin{tabular}{llcccc}
\toprule
$\varepsilon$ & & $T{=}1$ & $T{=}3$ & $T{=}6$ & $T{=}9$ \\
\midrule
\multirow{3}{*}{10} & AUC & $0.544$ & $0.573$ & $0.605$ & $0.594$ \\
 & $\sigma$ & 0.49 & 0.85 & 1.20 & 1.47 \\
 & utility & 15.7 & 24.9 & 34.8 & 47.7 \\
\midrule
\multirow{3}{*}{1} & AUC & $0.523$ & $0.557$ & $0.556$ & $0.575$ \\
 & $\sigma$ & 3.63 & 6.28 & 8.89 & 10.88 \\
 & utility & 12.3 & 18.7 & 32.0 & 39.0 \\
\bottomrule
\end{tabular}
\end{table}

\begin{table}[h]
\centering
\caption{AUC when attacking only the first $T$ rounds of PE run.}
\label{tab:abl-rounds-prefix}
\begin{tabular}{lllcccc}
\toprule
Modality & Dataset & $\varepsilon$ & $T{=}1$ & $T{=}2$ & $T{=}5$ & all rounds \\
\midrule
\multirow{4}{*}{Text} & \multirow{2}{*}{OpenReview} & $10$ & 0.600 & 0.619 & 0.655 & 0.687 \\
 & & $1$ & 0.519 & 0.532 & 0.537 & 0.543 \\
 & \multirow{2}{*}{MIMIC} & $10$ & 0.572 & 0.580 & 0.598 & 0.601 \\
 & & $1$ & 0.513 & 0.516 & 0.531 & 0.542 \\
\midrule
\multirow{3}{*}{Images} & CIFAR-10 & $10$ & 0.513 & 0.517 & 0.530 & 0.550 \\
 & Camelyon17 & $10$ & 0.516 & 0.527 & 0.533 & 0.555 \\
 & Cat & $10$ & 0.591 & 0.647 & 0.679 & 0.692 \\
\midrule
\multirow{2}{*}{Tabular} & \multirow{2}{*}{artificial-characters} & $10$ & 0.619 & 0.609 & 0.606 & 0.608 \\
 & & $1$ & 0.538 & 0.529 & 0.538 & 0.549 \\
\bottomrule
\end{tabular}
\end{table}

\paragraph{Privacy budget.}
Table~\ref{tab:abl-eps} shows the AUC on the tabular datasets across privacy budgets. Where the attack has signal, AUC drops as the privacy guarantee becomes stronger, e.g.~from $0.618$ without noise to $0.515$ at $\varepsilon=0.25$ on artificial-characters. On adult and person-activity, the AUC is close to $0.5$ for every budget, including $\varepsilon=\infty$, and the baseline attacks perform similarly. Hence, leakage is low even without DP noise. 

\begin{table}[h]
\centering
\caption{AUC across privacy budgets $\varepsilon$.}
\label{tab:abl-eps}
\begin{tabular}{lccccccc}
\toprule
Dataset & $\infty$ & 100 & 10 & 5 & 1 & 0.5 & 0.25 \\
\midrule
adult & 0.523 & 0.519 & 0.517 & 0.515 & 0.492 & 0.487 & 0.472 \\
artificial-characters & 0.618 & 0.597 & 0.588 & 0.574 & 0.531 & 0.533 & 0.515 \\
breast-cancer & 0.675 & 0.551 & 0.622 & 0.522 & 0.528 & 0.492 & 0.562 \\
person-activity & 0.512 & 0.509 & 0.494 & 0.499 & 0.489 & 0.504 & 0.484 \\
\bottomrule
\end{tabular}
\end{table}

\paragraph{Foundation model.}
Table~\ref{tab:abl-models} compares six generator LLMs on OpenReview generating $1{,}500$ synthetic samples each. The choice of model affects leakage much less than $\varepsilon$ and $T$. After $9$ rounds at $\varepsilon=10$, all results are between $0.578$ and $0.599$, with no trend associated with model size, and all follow the same pattern across rounds as Table~\ref{tab:abl-rounds-fixed}. 

\begin{table}[h]
\centering
\caption{AUC by generator LLM and number of rounds $T$ on OpenReview.}
\label{tab:abl-models}
\begin{tabular}{lcccc|cccc}
\toprule
 & \multicolumn{4}{c|}{$\varepsilon=10$} & \multicolumn{4}{c}{$\varepsilon=1$} \\
Model & $T{=}1$ & 3 & 6 & 9 & $T{=}1$ & 3 & 6 & 9 \\
\midrule
Llama-3.2-1B & 0.575 & 0.588 & 0.586 & 0.592 & 0.528 & 0.532 & 0.551 & 0.543 \\
Llama-3.2-3B & 0.560 & 0.577 & 0.586 & 0.587 & 0.546 & 0.552 & 0.553 & 0.543 \\
Llama-3.1-8B & 0.563 & 0.589 & 0.599 & 0.597 & 0.530 & 0.544 & 0.549 & 0.539 \\
Qwen2.5-1.5B & 0.543 & 0.568 & 0.600 & 0.592 & 0.516 & 0.535 & 0.543 & 0.539 \\
Qwen2.5-7B & 0.561 & 0.575 & 0.571 & 0.578 & 0.536 & 0.534 & 0.541 & 0.525 \\
Qwen2.5-14B & 0.555 & 0.578 & 0.588 & 0.599 & 0.515 & 0.530 & 0.543 & 0.532 \\
\bottomrule
\end{tabular}
\end{table}

\paragraph{Size of the private dataset.}
We subsample artificial-characters $|D|$ to $n\in\{500,\dots,7{,}152\}$ instances and run PE with fixed $\sigma=1.84$ (Table~\ref{tab:abl-n}). Here, $n$ is the number of private instances that vote in PE, and $D_{target}$ consists of the 250 sampled members from the train split of size $|n|$ and the same 213 non-members. When using a reference dataset, EVO-MIA is largely unaffected by $n$. In contrast, the AUC of EVO-MIA without the auxiliary set $\mathrm{Ref}$, i.e.~assuming every candidate is equally likely to receive a vote, drops from $0.636$ to $0.564$. 

\begin{table}[h]
\centering
\caption{AUC by private dataset size $n$ on artificial-characters.}
\label{tab:abl-n}
\begin{tabular}{lcccccc}
\toprule
$n$ & 500 & 1{,}000 & 2{,}000 & 3{,}500 & 5{,}000 & 7{,}152 \\
\midrule
EVO-MIA & 0.653 & 0.642 & 0.653 & 0.657 & 0.643 & 0.648 \\
EVO-MIA without $\mathrm{Ref}$& 0.636 & 0.591 & 0.568 & 0.573 & 0.567 & 0.564 \\
\bottomrule
\end{tabular}
\end{table}

\end{document}

%% file: diagram_overview.tex
\begin{figure}[t] 
\centering
\colorlet{privcol}{red!70}
\colorlet{synthcol}{blue!60}
\colorlet{votecol}{gray!80}
\colorlet{memcol}{green!60!black}
\colorlet{nonmemcol}{orange!90!black}
\resizebox{\columnwidth}{!}{%
\begin{tikzpicture}[
  >={Stealth[length=2mm]},
  record/.style={circle,draw=privcol!85,fill=privcol!25,minimum size=4.5pt,inner sep=0pt},
  target/.style={circle,draw=privcol,fill=privcol!55,minimum size=6.5pt,inner sep=0pt,line width=0.8pt},
  synth/.style={circle,draw=synthcol!85,fill=synthcol!30,minimum size=7pt,inner sep=0pt},
  synthlab/.style={synthcol!70!black,font=\footnotesize,inner sep=1.5pt},
  binlab/.style={synthcol!70!black,font=\footnotesize,anchor=north},
  vote/.style={->,votecol,semithick,shorten >=1.5pt,shorten <=1pt},
  stage/.style={->,line width=1.1pt,gray!60},
  tag/.style={rounded corners=2.5pt,draw,line width=0.9pt,font=\footnotesize,inner sep=3.5pt},
]
\begin{scope}
  \node[synth,label={[synthlab]above right:$a$}] (s1) at (0.75,0.45) {};
  \node[synth,label={[synthlab]above left:$c$}]  (s2) at (1.85,0.95) {};
  \node[synth,label={[synthlab]above right:$b$}] (s3) at (1.55,-0.45) {};
  \node[record] (r1) at (0.05,0.85) {};
  \node[record] (r2) at (0.10,0.05) {};
  \node[record] (r3) at (2.55,1.15) {};
  \node[record] (r4) at (2.45,0.35) {};
  \node[record] (r5) at (2.25,-0.75) {};
  \node[record] (r6) at (0.95,-0.85) {};
  \draw[vote] (r1)--(s1); \draw[vote] (r2)--(s1);
  \draw[vote] (r3)--(s2); \draw[vote] (r4)--(s2);
  \draw[vote] (r5)--(s3); \draw[vote] (r6)--(s3);
\end{scope}
\draw[stage] (2.95,0.1) -- (3.75,0.1);
\begin{scope}[shift={(4.05,-0.55)}]
  \draw[gray!55] (0,0)--(1.75,0);
  \fill[synthcol!60] (0.15,0) rectangle ++(0.34,0.70);
  \fill[synthcol!60] (0.70,0) rectangle ++(0.34,1.05);
  \fill[synthcol!60] (1.25,0) rectangle ++(0.34,0.70);
  \node[binlab] at (0.32,-0.10) {$a$};
  \node[binlab] at (0.87,-0.10) {$b$};
  \node[binlab] at (1.42,-0.10) {$c$};
\end{scope}
\draw[stage] (6.05,0.1) -- (6.85,0.1);
\begin{scope}[shift={(7.15,-0.55)}]
  \node[target,label={[privcol,font=\footnotesize]above:$x^\star$}] (tgt) at (-0.35,1.35) {};
  \draw[gray!55] (0,0)--(1.75,0);
  \fill[synthcol!60] (0.15,0) rectangle ++(0.34,0.70);
  \fill[synthcol!30] (0.70,0) rectangle ++(0.34,1.05);   
  \fill[synthcol!60] (1.25,0) rectangle ++(0.34,0.70);
  \fill[privcol!45] (0.70,1.05) rectangle ++(0.34,0.30);
  \draw[privcol,line width=0.9pt] (0.70,1.05)--++(0.34,0);
  \draw[vote,privcol,dashed,shorten >=2pt] (tgt) .. controls +(down:0.5) and +(up:0.5) .. (0.87,1.40);
  \node[binlab] at (0.32,-0.10) {$a$};
  \node[binlab] at (0.87,-0.10) {$b$};
  \node[binlab] at (1.42,-0.10) {$c$};
\end{scope}
\node[tag,memcol,draw=memcol,fill=memcol!8] (mem) at (10.75,0.6) {member};
\node[tag,nonmemcol,draw=nonmemcol] (non) at (10.75,-0.4) {non-member};
\draw[->,memcol,line width=1pt] (9.05,0.1) .. controls +(right:0.5) and +(left:0.5) .. (mem.west);
\draw[->,nonmemcol,line width=0.9pt,dashed] (9.05,0.1) .. controls +(right:0.5) and +(left:0.5) .. (non.west);
\node[font=\small] at (1.30,-1.45) {Voting};
\node[font=\small] at (4.92,-1.45) {Histogram};
\node[font=\small] at (8.02,-1.45) {Membership test};
\end{tikzpicture}%
} 
\caption{Overview of the voting mechanism and the proposed MIA. Private instances cast votes for the synthetic candidates. A noisy vote histogram over these candidates is released. The adversary tests if the target instance $x^\star$ nudges the histogram, ultimately classifying it as a member or non-member.}
\label{fig:voting_mia}
\end{figure}
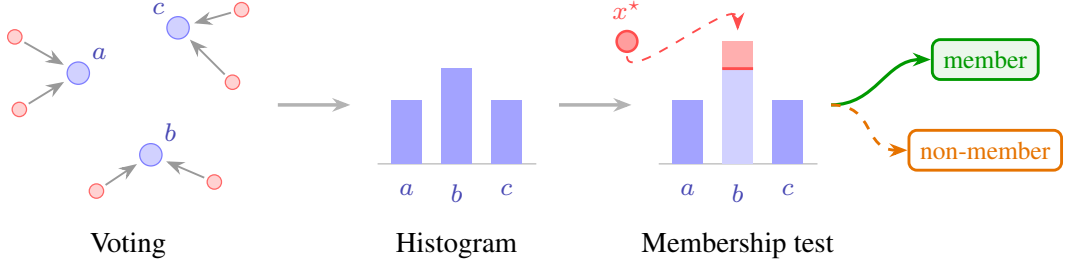

%% file: diagram_setup.tex
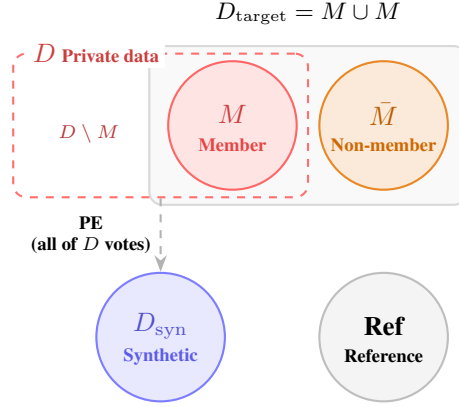
\begin{figure}[!h]
\centering
\colorlet{privcol}{red!70}
\colorlet{synthcol}{blue!60}
\colorlet{nonmemcol}{orange!90!black}
\colorlet{refcol}{gray!60}

\begin{tikzpicture}[
  font=\small,
  every node/.style={align=center},
  >={Stealth[length=2.2mm]},
  softbox/.style={draw=gray!30, thick, rounded corners=5pt, fill=gray!5},
  privbox/.style={draw=privcol!80, thick, dashed, rounded corners=5pt},
  memcircle/.style={circle, draw=privcol!85, fill=privcol!15, thick, minimum size=1.7cm, inner sep=0pt},
  nonmemcircle/.style={circle, draw=nonmemcol!85, fill=nonmemcol!15, thick, minimum size=1.7cm, inner sep=0pt},
  syncircle/.style={circle, draw=synthcol!85, fill=synthcol!15, thick, minimum size=1.7cm, inner sep=0pt},
  refcircle/.style={circle, draw=refcol!85, fill=refcol!15, thick, minimum size=1.7cm, inner sep=0pt}
]

\node[softbox, minimum width=4.2cm, minimum height=2.1cm] (Dbox) at (1.0, 0) {};
\node[above=1.5mm of Dbox.north, font=\small\bfseries] {$D_{\mathrm{target}} = M \cup \bar{M}$};

\node[privbox, minimum width=3.9cm, minimum height=1.9cm] (Dpriv) at (-0.95, 0) {};
\node[fill=white, inner sep=2pt, anchor=west, text=privcol!80!black]
  at ($(Dpriv.north west)+(0.2,0)$)
  {{\normalsize\bfseries $D$}\ {\scriptsize\bfseries Private data}};
\node[font=\scriptsize, text=privcol!70!black] at (-1.9, -0.1) {$D \setminus M$};

\node[memcircle] (M) at (0, 0) {};
\node[font=\normalsize\bfseries, text=privcol!80!black] at (0, 0.15) {$M$};
\node[font=\scriptsize\bfseries, text=privcol!90!black] at (0, -0.25) {Member};

\node[nonmemcircle] (Mbar) at (2.0, 0) {};
\node[font=\normalsize\bfseries, text=nonmemcol!80!black] at (2.0, 0.15) {$\bar{M}$};
\node[font=\scriptsize\bfseries, text=nonmemcol!90!black] at (2.0, -0.25) {Non-member};

\node[syncircle] (Dsyn) at (-0.95, -2.8) {};
\node[font=\normalsize\bfseries, text=synthcol!80!black] at (-0.95, -2.65) {$D_{\mathrm{syn}}$};
\node[font=\scriptsize\bfseries, text=synthcol!90!black] at (-0.95, -3.05) {Synthetic};

\draw[->, thick, dashed, draw=gray!60] (Dpriv.south) --
  node[left, font=\scriptsize\bfseries] {PE\\(all of $D$ votes)} (Dsyn.north);

\node[refcircle] (Ref) at (2.0, -2.8) {};
\node[font=\normalsize\bfseries] at (2.0, -2.65) {Ref};
\node[font=\scriptsize\bfseries] at (2.0, -3.05) {Reference};

\end{tikzpicture}
\caption{The target dataset $D_{\mathrm{target}} = M \cup \bar{M}$ consists of the
members $M \subseteq D$, a subset of private instances used by PE to generate $D_{\mathrm{syn}}$, and a
disjoint set of non-members $\bar{M}$ from the same distribution. An
independent reference pool $\mathrm{Ref}$ is used during the attack.}
\label{fig:data_universe}
\end{figure}

%% file: references.bib
@inproceedings{van2023membership,
  title={Membership Inference Attacks against Synthetic Data through Overfitting Detection},
  author={van Breugel, Boris and Sun, Hao and Qian, Zhaozhi and van der Schaar, Mihaela},
  booktitle={International Conference on Artificial Intelligence and Statistics},
  pages={3493--3514},
  year={2023},
  organization={PMLR}
}

@inproceedings{trandifferentially2026,
  title={Differentially Private Synthetic Data via APIs 4: Tabular Data},
  author={Tran, Toan and Backurs, Arturs and Lin, Zinan and Reis, Victor and Xiong, Li and Yekhanin, Sergey},
  booktitle={Forty-Third International Conference on Machine Learning},
  year={2026}
}

@inproceedings{
qu2025tabicl,
title={Tab{ICL}: A Tabular Foundation Model for In-Context Learning on Large Data},
author={Jingang QU and David Holzm{\"u}ller and Ga{\"e}l Varoquaux and Marine Le Morvan},
booktitle={Forty-Second International Conference on Machine Learning},
year={2025},
}

@inproceedings{chen2020gan,
  title={Gan-leaks: A taxonomy of membership inference attacks against generative models},
  author={Chen, Dingfan and Yu, Ning and Zhang, Yang and Fritz, Mario},
  booktitle={Proceedings of the 2020 ACM SIGSAC conference on computer and communications security},
  pages={343--362},
  year={2020}
}

@article{sun2025synbench,
  title={SynBench: A Benchmark for Differentially Private Text Generation},
  author={Sun, Yidan and Schlegel, Viktor and Nandakumar, Srinivasan and Zahid, Iqra and Wu, Yuping and Wu, Yulong and Li, Hao and Zhang, Jie and Del-Pinto, Warren and Nenadic, Goran and others},
  journal={arXiv preprint arXiv:2509.14594},
  year={2025}
}

@inproceedings{lin2024differentially,
	title        = {{Differentially Private Synthetic Data via Foundation Model {API}s 1: Images}},
	author       = {Zinan Lin and Sivakanth Gopi and Janardhan Kulkarni and Harsha Nori and Sergey Yekhanin},
	year         = 2024,
	booktitle    = {{The Twelfth International Conference on Learning Representations}},
}

@article{ozturk2026towards,
  title={{Towards Useful and Private Synthetic Omics: Community Benchmarking of Generative Models for Transcriptomics Data}},
  author={{\"O}zt{\"u}rk, Hakime and Afonja, Tejumade and J{\"a}lk{\"o}, Joonas and Binkyte, Ruta and Rodriguez-Mier, Pablo and Lobentanzer, Sebastian and Wicks, Andrew and Kreuer, Jules and Ouaari, Sofiane and Pfeifer, Nico and others},
  journal={bioRxiv},
  pages={2026--03},
  year={2026},
  publisher={{Cold Spring Harbor Laboratory}}
}

@article{eckardt2023mimicking,
	title        = {{Mimicking Clinical Trials with Synthetic {Acute} {Myeloid} {Leukemia} Patients Using Generative Artificial Intelligence}},
	author       = {Eckardt, Jan-Niklas and Hahn, Waldemar and R{\"o}llig, Christoph and Stasik, Sebastian and Platzbecker, Uwe and M{\"u}ller-Tidow, Carsten and Serve, Hubert and Baldus, Claudia D and Schliemann, Christoph and Schaefer-Eckart, Kerstin and others},
	year         = 2023,
	journal      = {{Blood}},
	publisher    = {Elsevier},
	volume       = 142,
	pages        = 2268
}

@inproceedings{hou2024pre,
	title        = {{{PrE-Text}: Training Language Models on Private Federated Data in the Age of {LLMs}}},
	author       = {Hou, Charlie and Shrivastava, Akshat and Zhan, Hongyuan and Conway, Rylan and Le, Trang and Sagar, Adithya and Fanti, Giulia and Lazar, Daniel},
	year         = 2024,
	booktitle    = {{International Conference on Machine Learning}},
}

@article{chien2026maple,
  title={{MAPLE: Metadata Augmented Private Language Evolution}},
  author={Chien, Eli and Hu, Yuzheng and McKenna, Ryan and Wu, Shanshan and Xu, Zheng and Kairouz, Peter},
  journal={arXiv preprint arXiv:2603.19258},
  year={2026}
}

@inproceedings{sun2025evaluating,
	title        = {{Evaluating Differentially Private Generation of Domain-Specific Text}},
	author       = {Sun, Yidan and Schlegel, Viktor and Kolumam Nandakumar, Srinivasan and Zahid, Iqra and Wu, Yuping and Del-Pinto, Warren and Nenadic, Goran and Siew Kei, Lam and Zhang, Jie and Bharath, Anil},
	year         = 2025,
	booktitle    = {{Proceedings of the 34th ACM International Conference on Information and Knowledge Management}},
	pages        = {5273--5278}
}

@inproceedings{xiedifferentially,
	title        = {{Differentially Private Synthetic Data via Foundation Model {API}s 2: Text}},
	author       = {Xie, Chulin and Lin, Zinan and Backurs, Arturs and Gopi, Sivakanth and Yu, Da and Inan, Huseyin A. and Nori, Harsha and Jiang, Haotian and Zhang, Huishuai and Lee, Yin Tat and Li, Bo and Yekhanin, Sergey},
	year         = 2024,
	booktitle    = {{International Conference on Machine Learning}},
}

@inproceedings{auditing2023tight,
  title={Tight auditing of differentially private machine learning},
  author={Nasr, Milad and Hayes, Jamie and Steinke, Thomas and Balle, Borja and Tram{\`e}r, Florian and Jagielski, Matthew and Carlini, Nicholas and Terzis, Andreas},
  booktitle={32nd USENIX Security Symposium (USENIX Security 23)},
  pages={1631--1648},
  year={2023}
}

@article{tran2026automated,
  title={Automated Membership Inference Attacks (AutoMIA): Discovering MIA Signal Computations using LLM Agents},
  author={Tran, Toan and Kotevska, Olivera and Xiong, Li},
  journal={Transactions on Machine Learning Research},
  year={2026}
}

@article{wang2026synthesize,
  title={Synthesize privacy-preserving high-resolution images via private textual intermediaries},
  author={Wang, Haoxiang and Lin, Zinan and Yu, Da and Zhang, Huishuai},
  journal={Advances in Neural Information Processing Systems},
  volume={38},
  pages={39557--39589},
  year={2026}
}

@inproceedings{swanberg2025api,
  title     = {Is API Access to LLMs Useful for Generating Private Synthetic Tabular Data?},
  author    = {Swanberg, Marika and McKenna, Ryan and Roth, Edo and Cheu, Albert and Kairouz, Peter},
  booktitle = {ICLR 2025 Workshop on Synthetic Data for Machine Learning (SynthData)},
  year      = {2025},
}

@inproceedings{
zou2026pesgd,
title={{PE}-{SGD}: Differentially Private Deep Learning via Evolution of Gradient Subspace for Text},
author={Tianyuan Zou and Zinan Lin and Sivakanth Gopi and Yang Liu and Ya-Qin Zhang and Robert Sim and Xin Deng and Sergey Yekhanin},
booktitle={Fourteenth International Conference on Learning Representations},
year={2026},
}

@inproceedings{zou2025contrastive,
  title={Contrastive Private Data Synthesis via Weighted Multi-PLM Fusion},
  author={Zou, Tianyuan and Liu, Yang and Li, Peng and Xiong, Yufei and Zhang, Jianqing and Liu, Jingjing and Ye, Xiaozhou and Ouyang, Ye and Zhang, Ya-Qin},
  booktitle={International Conference on Machine Learning},
  pages={80846--80872},
  year={2025},
  organization={PMLR}
}

@inproceedings{zhangpcevolve2025,
  title={PCEvolve: Private Contrastive Evolution for Synthetic Dataset Generation via Few-Shot Private Data and Generative APIs},
  author={Zhang, Jianqing and Liu, Yang and FU, JIE and Hua, Yang and Zou, Tianyuan and Cao, Jian and Yang, Qiang},
  booktitle={Forty-second International Conference on Machine Learning},
  year={2025}
}

@inproceedings{shokri2017membership,
	title        = {{Membership Inference Attacks Against Machine Learning Models}},
	author       = {Shokri, Reza and Stronati, Marco and Song, Congzheng and Shmatikov, Vitaly},
	year         = 2017,
	booktitle    = {{2017 IEEE Symposium on Security and Privacy (SP)}},
	pages        = {3--18},
	organization = {IEEE}
}

@inproceedings{hu2023sok,
	title        = {{{SoK}: Privacy-Preserving Data Synthesis}},
	author       = {Hu, Yuzheng and Wu, Fan and Li, Qinbin and Long, Yunhui and Garrido, Gonzalo and Ge, Chang and Ding, Bolin and Forsyth, David and Li, Bo and Song, Dawn},
	year         = 2024,
	booktitle    = {{IEEE Symposium on Security and Privacy (SP)}},
	pages        = {4696--4713}
}

@inproceedings{golob2025privacy,
	title        = {{Privacy vulnerabilities in marginals-based synthetic data}},
	author       = {Golob, Steven and Pentyala, Sikha and Maratkhan, Anuar and De Cock, Martine},
	year         = 2025,
	booktitle    = {{2025 IEEE Conference on Secure and Trustworthy Machine Learning (SaTML)}},
	pages        = {977--995},
	organization = {IEEE}
}

@inproceedings{menziessynthetic__camda,
	title        = {{Synthetic Data Generation for bulk {RNA}-seq Data: A {CAMDA} Health Challenge Analysis}},
	author       = {Menzies, Shane and Filienko, Daniil and Golob, Steven and Pentyala, Sikha and Banerjee, Jineta and Foschini, Luca and de Cock, Martine},
	year         = 2025,
	booktitle    = {{Annual International Conference on Critical Assessment of Massive Data Analysis (CAMDA), collocated with ISMB/ECCB}}
}

@article{hayes2017logan,
  title={{LOGAN}: Membership inference attacks against generative models},
  author={Hayes, Jamie and Melis, Luca and Danezis, George and De Cristofaro, Emiliano},
  journal={Proceedings on Privacy Enhancing Technologies},
  volume={1},
  pages={133–152},
  year={2019}
}
